\documentclass[useAMS,usenatbib]{mnras}
\usepackage{amssymb}
\usepackage{amsfonts}
\usepackage{tabularx}
\usepackage{graphicx}
\usepackage{ulem}
\usepackage{array}
\usepackage{color}
\usepackage{booktabs}
\usepackage[fleqn]{amsmath}
\makeatletter
\newcommand*{\rom}[1]{\expandafter\@slowromancap\romannumeral #1@}
\makeatother
\numberwithin{equation}{section}
\def\d{{\rm d}}

\newcommand{\p}{\partial}

\newcommand{\bv}{{\mbox{\boldmath $v$}}}

\newcommand{\br}{{\mbox{\boldmath $r$}}}

\begin{document}

\title[]{Simulation of the loss-cone instability in spherical systems. II. Dominating Keplerian potential}
\author[E.\,V.\,Polyachenko et al.]
   { E.~V.~Polyachenko\,$^1$\thanks{E-mail: epolyach@inasan.ru},
     P.~Berczik\,$^{2,3,4}$\thanks{E-mail: berczik@mao.kiev.ua},
     A.~Just\,$^3$\thanks{E-mail: just@ari.uni-heidelberg.de},
     I.~G.~Shukhman\,$^5$\thanks{E-mail: shukhman@iszf.irk.ru}\\
     $^1 ${\it\small Institute of Astronomy, Russian Academy of Sciences,}
          {\it\small 48 Pyatnitskya St., Moscow 119017, Russia}\\
     $^2 ${\it\small The International Center of Future Science of the Jilin University, 2699 Qianjin St., 130012 Changchun City, PR China }\\
     $^3 ${\it\small Zentrum f\"ur Astronomie der Universit\"at Heidelberg, Astronomisches Rechen-Institut,  }
          {\it\small M\"{o}nchhofstr. 12-14, 69120 Heidelberg, Germany} \\
     $^4 ${\it\small Main Astronomical Observatory, National Academy of Sciences of Ukraine, MAO/NASU,}
          {\it\small 27 Akad. Zabolotnoho St. 03680 Kyiv, Ukraine} \\
     $^5 ${\it\small Institute of Solar-Terrestrial Physics, Russian Academy of Sciences,}
          {\it\small Siberian Branch, P.O. Box 291, Irkutsk 664033, Russia }}



\maketitle

\begin{abstract}
A new so-called `gravitational loss-cone instability' in stellar systems has recently been investigated theoretically in the framework of linear perturbation theory and proved to be potentially important in understanding the physical processes in centres of galaxies, star clusters, and the Oort comet cloud. Using N-body simulations, we confirm previous findings and go beyond the linear theory. Unlike the well-known instabilities, the new one shows no notable change in spherical geometry of the cluster, but it significantly accelerates the speed of diffusion of particles in phase space leading to a repopulation of the loss cone and early instability saturation.
\end{abstract}


\section{Introduction}
\label{sec:intro}%

In the pioneer paper, \citet{1991SvAL...17..371P} studied a simple analytical model of a low-mass stellar disc in a dominating point-mass potential. He argued that if the radial-orbit instability \citep[for review, see][]{2015MNRAS.451..601P} is suppressed, a distant relative of the loss-cone instability in plasma~\citep{1965PhFl....8..547R} still may occur.

\citet{2005ApJ...625..143T} examined spherical and disc systems and found that stability properties are determined by the dependence of the distribution function (DF) $F$ on angular momentum $L$. In particular, flattened, nonrotating systems were found to be secularly stable if $\p F/\p L < 0$ at constant energy, while for $\p F/\p L > 0$ they are generally unstable. Based on the analysis of dipole and quadrupole distortions, the spherical systems in which $F=0$ at $L=0$ (an empty loss cone) were claimed to be stable for any sign of $\p F/\p L$.

Using a newly elaborated matrix method for spherical systems analogous to the matrix method for discs~\citep{EP05}, \citet{2007MNRAS.379..573P} found that the instability in spherical systems considered by Tremaine is nevertheless possible, but for the higher-order distortions starting from octupole. In the subsequent paper \citep[][hereafter Paper I]{2008MNRAS.386.1966P}, we refined our claim that it is only relevant to non-monotonic dependence of the DF on $L$. In analogy with plasma, we called it gravitational loss-cone instability (gLCI).

These analytical studies assume that a cluster of mass $M_*$ is embedded in the dominating potential of a central point mass $M$, so that $\varepsilon\equiv M_*/M$ is a small parameter. The gravitational potential
\begin{equation}
    \Phi(r) = -\frac{GM}r + \Phi_*(r)\,,
\end{equation}
consists of the dominant Keplerian part, and $\Phi_*$ due to selfgravity of the cluster. If the latter is neglected, a particle travels along the fixed ellipses with semiaxes $a = GM/(2|E|)$, $b=L/(2|E|)^{1/2}$ and eccentricity $e = (1-\alpha^2)^{1/2}$, where $\alpha\equiv L/L_{\rm circ}$, $L_{\rm circ} = GM/(2|E|)^{1/2}$ is the angular momentum of a particle with energy $E$ on a circular orbit. The radial and orbital frequencies are equal and independent of $L$: $\Omega_1 = \Omega_2 = (2|E|)^{3/2}/(GM)$.

Accounting for the selfgravity leads to slow orbit precession, which is always retrograde \citep{2007MNRAS.379..573P}. Generally, there are four characteristic timescales in such a system~\citep{2005ApJ...625..143T}: dynamical time $t_{\rm dyn}\sim \Omega_1^{-1}$, precession time $t_{\rm pr}\sim \varepsilon^{-1} t_{\rm dyn}$, resonant relaxation time $t_{\rm res}\sim N\varepsilon^{-1} t_{\rm dyn}$~\citep{1996NewA....1..149R}, where $N$ is the number of particles, and two-body relaxation time $t_{\rm relax}\sim N\varepsilon^{-2} t_{\rm dyn}$. For $N \gg 1$ these timescales are well separated:
\begin{equation}
    t_{\rm dyn} \ll t_{\rm pr} \ll t_{\rm res} \ll t_{\rm relax}\,.
\end{equation}
If the system is dynamically stable, its evolution is driven by relaxation on timescales $t_{\rm res}$ or $t_{\rm relax}$. However, if there is an instability due to collective mechanism in the distribution of orbits, its timescale $t_{\rm ins}$ is expected to be close to the orbital movement time, i.e. $t_{\rm pr}$. As found for several models in Paper I, the instability time is in fact only a few times larger than the precession time. Thus instability becomes the main change driving process overtaking both kinds of relaxation. However, the instability is slow in the dynamical timescale and requires by a factor of $\varepsilon^{-1}$ longer simulations than standard simulations of systems in which the dynamical time is largely determined by selfgravity.

This paper is aimed to simulate numerically one of the models studied earlier theoretically using linear perturbation theory (Paper I). It serves two purposes: to confirm our theoretical finding and to go beyond the linear approximation and study the consequences of the instability development. In a companion paper~\citep{P19h}, we simulated the instability in a dominating harmonic potential and showed that unlike, e.g., the well-known bar instability in discs, gLCI did not change the shape of the cluster, but led to the enhanced diffusion resulting in the repopulation of the loss cone and cease of the instability.

In addition to the peculiarities of the harmonic case, the dominating central point mass requires additional care for particles close to the centre. It is worthwhile to emphasize that popular N-body codes incorporating the black hole dynamics or the external central mass potential~\citep[e.g., Nbody6++,][]{WSANBKN2015} are proved to be effective in the opposite limit $M \ll M_*$, but not in the case of interest here. Thus additional investigations were needed to improve the available numerical schemes for the current task.

The paper is organized as follows. The model and the set-up are described in Section 2, along with details of N-body simulations. Results containing a comparison of the instability growth rate with the linear perturbation theory, and analysis of peculiarities of the cluster evolution are given in Section 3. The final section 4 contains conclusions, final remarks, and describes future perspectives.

\section{The model, N-body set-up and simulations}
\label{sec:nbody-set}

\subsection{The model}

For numerical simulations, we choose a cluster in which all particles have the same energy $E_0$, i.e. the DF is $F(E,L) = A \delta(E-E_0)\,f(\alpha)$, where $\delta(x)$ is the Dirac delta-function, $A = M_* \Omega_1/(16\pi^3 L_{\rm circ}^2)$ is a normalization constant. The stability properties are determined by the angular momentum dependence, which is assumed to be non-monotonic:
\begin{equation}
    f(\alpha) = \frac{N_n}{\alpha_T^2} x^n {\rm e}^{-x}\,,\quad x\equiv \frac{\alpha^2}{\alpha_T^2}\,.
    \label{eq:df}
\end{equation}
Here $E=v^2/2+\Phi(r)$  and $L=|\bv\times\br|$ are (specific) energy and angular momentum of a particle; $N_n$ is the normalization constant to fulfil the condition $\int_0^1 \d \alpha\,\alpha f(\alpha) = 1$. There are two model parameters $\alpha_T$ and $n$: the former controls the fraction of nearly radial orbits, while the second one controls the width of the loss cone. We define the potential $\Phi_*(r)$ to be equail to $-GM_*/R$ at the edge of the sphere, so $E_0 = -G(M+M_*)/R$.

\begin{figure}
\centering
  \centerline{\includegraphics[width=\linewidth]{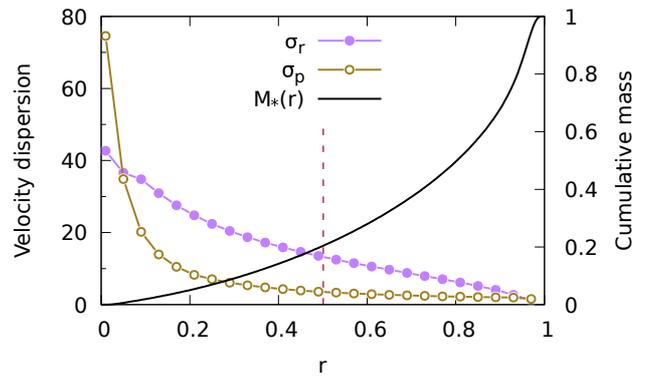} }
  \caption{The model for $M=100$: velocity dispersion profiles for radial and transversal directions and cumulative mass $M_*(r)$. The vertical dashed line shows the radius of the circular orbit $r_{\rm circ}$.}
  \label{fig:model}
\end{figure}

For the modelling, we adopt units in which $G = M_* = R = 1$. The model parameters were fixed ($\alpha_T = 0.173$, $n=3$), while the central mass $M$ and the number of particles $N$ varied. For $M=100$, the cumulative mass and the velocity dispersion profiles $\sigma_r$ and $\sigma_p$ ($=\sigma_\theta=\sigma_\varphi$) are given in Fig.\,\ref{fig:model}.

\begin{figure}
\centering
  \centerline{\includegraphics[width=\linewidth]{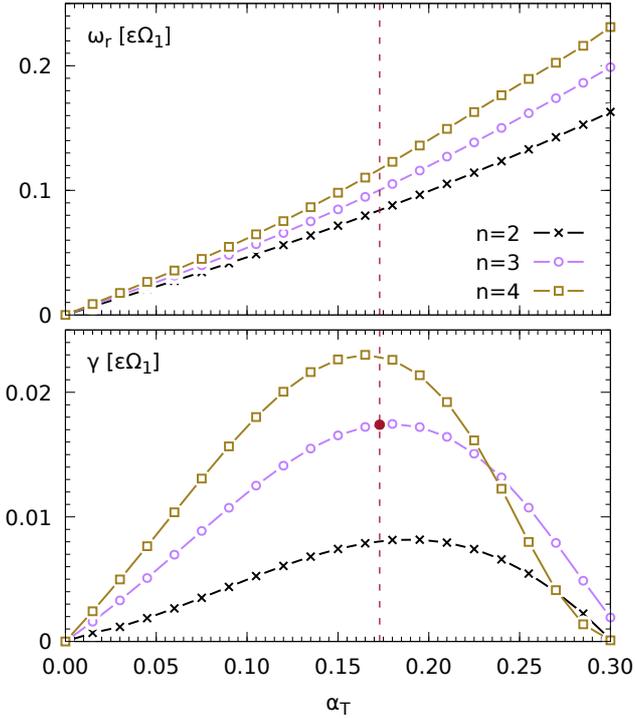}}
  \caption{Octupole real and imaginary parts of the frequency $\omega = \omega_r + i\gamma$ in the slow units $\varepsilon\Omega_1$ v.s. parameter $\alpha_T$ for several values of $n$. The vertical dashed line shows $\alpha_T=0.173$; the filled circle marks the near maximum growth rate of the curve $n=3$ we set out to reproduce.}
  \label{fig:omega_ri}
\end{figure}

For the family of Eq.\,(\ref{eq:df}), linear perturbation theory predicts the absence of instability for dipole and quadrupole perturbations, and the complex frequency of the octupole perturbations shown in Fig.\,\ref{fig:omega_ri}. Positive $\gamma$ means exponential growth of the this harmonic in time. In particular, the values of $\alpha_T$ and $n$, adopted for the simulations, correspond to $\omega_r=0.1\,\varepsilon\Omega_1$, $\gamma=0.0174\,\varepsilon\Omega_1$.

\subsection{Initial distribution}

The initial self-consistent model was constructed iteratively using a relation between the density and the DF:
\begin{equation}
    r\rho(r) = 4\pi A \int\limits_0^{L_{\rm max}(r)} \frac{f(\alpha) L\d L}{[L_{\rm max}^2(r) - L^2]^{1/2}}\,,
    \label{eq:rrho}
\end{equation}
where $L_{\rm max}(r)$ is determined from equation
\begin{equation}
    L^2_{\rm max}(r) = 2r^2(E_0 - \Phi(r))\,,
\end{equation}
and $L_{\rm circ}$ is determined by the radius of the circular orbit $r_{\rm circ}$,
\begin{align}
    &2E_0 - 2\Phi(r_{\rm circ})-r_{\rm circ}\Phi'(r_{\rm circ}) = 0\,, \\
    &L^2_{\rm circ} = r^3_{\rm circ}\Phi'(r_{\rm circ})\,.
\end{align}
To avoid the singularity at $L=L_{\rm max}$, it is helpful to introduce a new variable $s$ by $\alpha = (1-s^2)^{1/2} L_{\rm max}(r)/L_{\rm circ}$. Eq.\,(\ref{eq:rrho}) then reads:
\begin{equation}
    r\rho(r) = 4\pi A L_{\rm max}(r) \int\limits_0^1 f\bigl(\alpha(s)\bigr) \d s\,.
    \label{eq:rrho1}
\end{equation}
As zero-order approximation, $\Phi_*$ was set to zero, and an approximation was obtained for $\rho(r)$. The normalization constant $A$ was recalculated after each iteration to provide the total mass $M_*$ for the cluster. The next approximation for the potential $\Phi_*$ was obtained from a numerical solution of the Poisson equation
\begin{equation}
    \frac{1}{r^2} \frac{\d}{\d r} r^2 \frac{\d }{\d r} \Phi_*(r) = 4\pi G \rho(r)\,.
    \label{eq:poi}
\end{equation}

After obtaining self-consistent profiles $\rho(r)$ and $\Phi_*(r)$, one can build a random realization of the DF. First, we find $L$ of the particle using the standard von Neumann rejection technique for Eq.\,(\ref{eq:df}). Energy $E=E_0$ and angular momentum $L$ determine the orbit. Second, we set the radius of the particle, which is uniformly distributed in the radial angle variable, $w$. The dependence $r(w; E,L)$ can be obtained directly by orbit integration, if the number of particles is not too large, or interpolating between reference orbits from a library. The radius $r$ immediately gives the square of radial velocity $v_r^2$, and the absolute value of transversal velocity $v_\perp = (v_\theta^2 + v_\varphi^2)^{1/2}= L/r$. The velocity components $(v_\theta, v_\varphi)$  are uniformly distributed over a polar angle $\psi$ in the velocity polar coordinates $(v_{\perp},\psi)$. Finally, the spatial coordinates are obtained as a uniform distribution on a sphere.

For our default model with $N~=~10^5$ and the central point mass $M~=~100$, this procedure gives a nearly spherically symmetric N-body realisation with center-of-mass offset $~1.5\cdot 10^{-3}$. Due to numerical noise, the initial amplitudes of the spherical harmonics $A_l$ (see Eq.\,(\ref{eq:A10})) for $1\leq l \leq 4$ are of the order of $10^{-3}$.



\subsection{N-body simulations}

The simulations presented in this work have been carried out
using the specially adapted direct N-body code $\varphi$-GPU
\citep{BNZ2011, BSW2013}. The basic concepts of the code are
based on the publicly available $\varphi$-GRAPE \citep{HGM2007}
code, an Aarseth N-body\,1-like code including an efficient
MPI parallelization and support for the special-purpose
hardware GRAPE \citep{FIMETSU1991} (currently Graphic Processing
Units - GPU). The current version of the $\varphi$-GPU code
has been fully rewritten in the {\sc C++} programming language
for the most effective use of GPUs.

The $\varphi$-GPU code is fully parallelized using the
MPI library. The MPI parallelization was done in the
same ``j'' particle parallelization scheme as in the
earlier $\varphi$-GRAPE code. All particles are
divided equally between the working nodes
(using the {\tt MPI\_Bcast()} command) and in each node
we calculate only the fractional forces for the particles
in the current time step, i.e. the so-called ``active''
or ``i'' particles. We get the full forces from all
particles acting on the active particles after the
global {\tt MPI\_Allreduce()} communication routine
is applied.

The code supports the use of individual timestep Hermite integration algorithms
of 4$^{th}$, 6$^{th}$ or 8$^{th}$ order. Furthermore, it includes
a hierarchical block time-step scheme. Compared to the earlier
$\varphi$-GRAPE \citep{HGM2007} implementation we obtain an additional
speed-up by a factor of $\times 2$ on the recent generation
of NVIDIA GPUs (i.e., K20 and V100 cards), depending on the
specific number of particles and computing nodes used.

Further details of this high-performance computing
code is be described in \citet{BNZ2011, BSW2013}.
The present code is well tested and already used to obtain important
results in our earlier large scale few million body simulations
~\citep{LLBS2017, LLBCS2012, WBSK2014}.

The $\varphi$-GPU code does not include the regularization \citep{MA1998}
of close encounters or binaries, so we use small softening to
avoid the formation of tight binaries during our simulation.
We use a typical Plummer-type softening between individual particles
($\epsilon = 10^{-4}$). The gravitational softening
for the interaction between particles and the central Keplerian
potential was set exactly equal to zero in the current set of runs.

We choose the conservative standard 4$^{th}$ order Hermite integration
parameter $\eta$~=~0.01. For the default run, the total
energy is conserved to a relative error $\approx$ 0.6 \% during
the full integration time of 1000 time units. 
The average random velocity of the cluster center of mass (CoM) was $\approx 2\cdot 10^{-4}$.

\section{Results}
\label{sec:nbody-res}

In order to compare results of our numerical simulations with theoretical predictions, we evaluate the spherical harmonics of the density distribution  
\begin{equation}
\rho(r,\theta,\varphi)=\sum\limits_{l=0}^{\infty} \rho_l(r,\theta,\varphi)\,,
\label{eq:rho}
\end{equation}
\begin{equation}
\rho_l(r,\theta,\varphi)=\sum\limits_{m=-l}^l C_l^m(r)\, Y_l^m(\theta,\phi)\,,
\end{equation}
where $Y_l^m(\theta,\phi)$ are fully normalised spherical functions, for each snapshot. The coefficients $C_l^m$ depend on radius and the orientation of the frame. In order to make a comparison simpler, we evaluate amplitudes
\begin{equation}
A_l = \frac1N \left[ \sum_{m=-l}^l \left| \sum\limits_{i=1}^N Y_l^{m} (\theta_i, \varphi_i)\right|^2 \right]^{1/2}
\end{equation}
invariant to the frame rotation~\citep[see Appendix \ref{app1} and][]{P19h}.

\begin{figure}
\centering
  \centerline{\includegraphics[width=\linewidth]{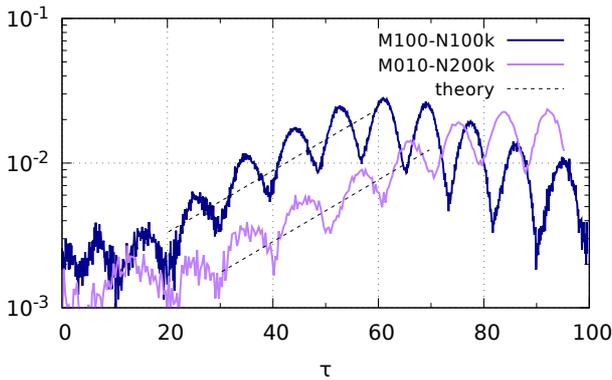}}
  \caption{Growth of the octupole perturbations $A_3$ in runs with $M=100$, $N=100$k, and $M=10$, $N=200$k v.s. instability slow time $\tau \equiv t/M^{1/2}$. The dashed lines shows the predicted theoretical growth.}
  \label{fig:a3abs}
\end{figure}

From Sec.\,\ref{sec:intro} and Sec.\,\ref{sec:nbody-set} we infer that the dynamical time scales with $M$ as $t_{\rm dyn} \sim M^{-1/2}$. Therefore, the instability time scales as $t_{\rm ins} \sim t_{\rm pr} \sim M^{1/2}$. Fig.\,\ref{fig:a3abs} compares the amplitudes of the octupole perturbations ($l=3$) for two runs in the instability time scale. We see the exponential growth of the amplitudes with the predicted slope, however the amplitudes are modulated by oscillations. These oscillations are due to {\it i}) the oscillation character of the eigenfrequency $\omega_r \ne 0$, and {\it ii}) the existense of another eigenfrequency with the same growth rate. Indeed, the integral equation (2.18) of Paper I has the form
$
\omega^2\psi_l={\hat L}_l\,\psi_l,
$
where ${\hat L}_l$
is the linear integral real operator. Hence, if $\omega^2$ is an eigen-value, then $(\omega^2)^*$ is also tan eigen-value of the problem. This means that if $\omega_r + i\gamma$ is a solution, then $-\omega_r + i\gamma$ also is a solution. A general growing solution then reads
\begin{equation}
F = \bigl[C_1 \exp(i \omega_r t) + C_2 \exp(-i \omega_r t)\bigr]\cdot\exp(\gamma t)\,,
\end{equation}
and
\begin{equation}
|F|^2 = \bigl[|C_1|^2 +\! |C_2|^2 + 2|C_1||C_2| \cos (2\omega_r\!+\!\phi)\bigr]\cdot
\exp(2\gamma t)\,,
\end{equation}
where $\phi$ is due to the difference of the complex phases of the amplitudes $C_{1,2}$. These oscillations  we see  in Fig.\,\ref{fig:a3abs}.

We also observe early saturation of the instability so that amplitudes remain below a level of 0.03, similar to the gLCI in a dominating harmonic potential~\citep{P19h}.  There, it was explained by rapid repopulation of the loss cone, i.e. the region of low angular momentum, due to diffusion enhanced by the instability.

For near-Keplerian models, we estimate the relaxation time using the standard formula for two-body relaxation
\begin{equation}
    t_{\rm relax} \sim \frac{N}{\ln N}\frac{R^{3/2} M^{3/2}}{G^{1/2} M^2_*},
 \end{equation}
 so that in dimensionless form the two-body relaxation time in units of the slow time $M^{1/2}$ is
 \begin{equation}
   \frac{t_{\rm relax}}{M^{1/2}}
     = \frac{N}{\ln N} \cdot  M = (2 - 9) \times 10^5,
     \label{eq:relax}
\end{equation}
where the value range covers the parameters shown Fig.\,\ref{fig:a3abs}. 

For the resonant relaxation~\citep{1996NewA....1..149R, 2005ApJ...625..143T}
\begin{equation}
    t_{\rm res}  \sim N\,\frac{R^{3/2} M^{1/2}}{G^{1/2} M_*}
\end{equation}
 so that in dimensionless form the resonance relaxation time in units of slow time $M^{1/2}$ is
  \begin{equation}
 \frac{ t_{\rm res}}{M^{1/2}}=N  \sim (1 - 2)\times 10^5\,.
    \label{eq:res}
\end{equation}
One can conclude that these two mechanisms alone cannot be responsible for a particle repopulation during the time of simulations.
\begin{figure*}
\centering
  \centerline{\includegraphics[width=\linewidth]{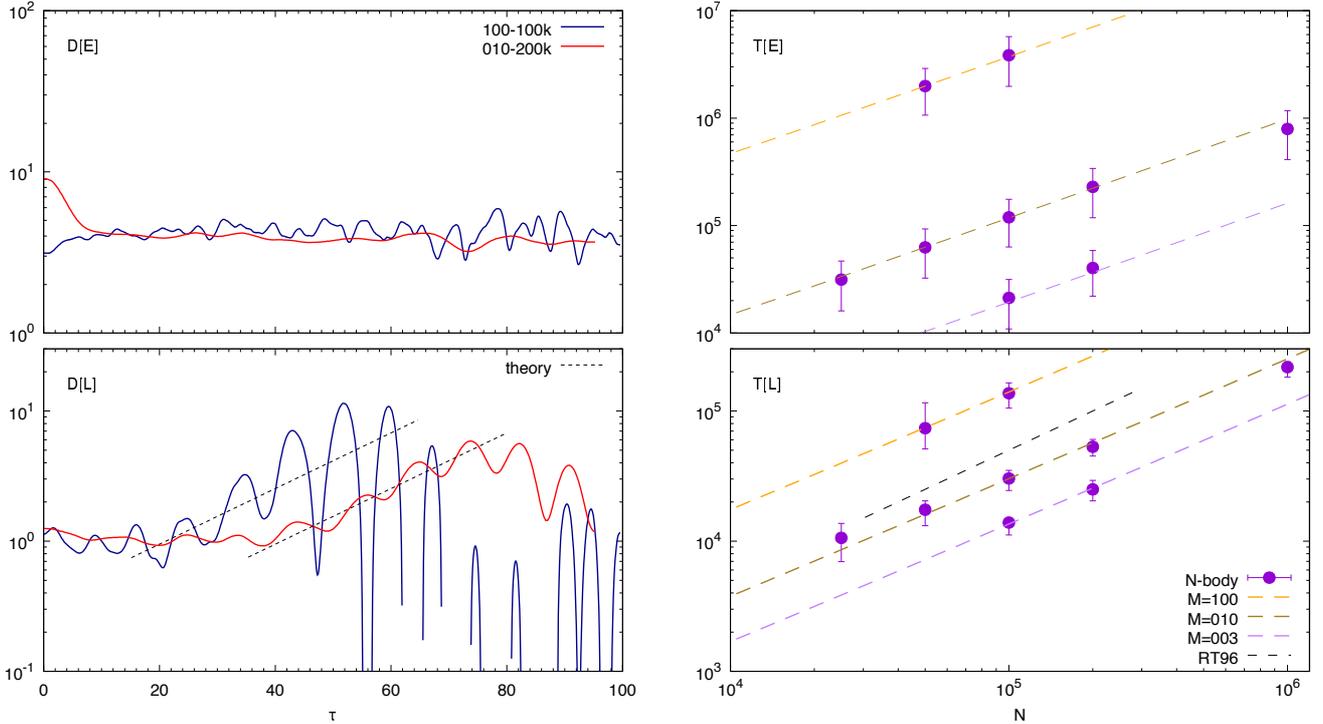}}
  \caption{Characteristics of the $E$- and $L$-diffusion (upper and lower panels, respectively). Left panels show normalized diffusion coefficients [see Eqs.\,(\ref{eq:D-E}) and (\ref{eq:D-L})] for runs $(M,N)$=(100, 100k), (10, 200k); black dashed lines in the lower left panel indicate slopes of the instability growth rate for the octupole harmonic from Fig.\,\ref{fig:a3abs}.  Right panels show the relaxation times [see Eqs.\,(\ref{eq:T-E}) and (\ref{eq:T-L})] for runs $M=3$, $N=$100k, 200k; $M=10$, $N=$25k, 50k, 100k, 200k, 1000k; $M=100$, $N=$50k, 100k. Colored dashed lines in the upper right panel show the time fit of Eq.\,(\ref{eq:relaxfit}) corresponding to two-body relaxation; in the lower right panel -- time fit (Eq.\,\ref{eq:resfit})) close to resonant relaxation; the black dashed line marked RT96 shows the theoretical expectation of \citet{1996NewA....1..149R}. Upper and lower values of errorbars show maximum and minimum times obtained from variations of the diffusion coefficients, while the filled circles give average values.}
  \label{fig:DE}
\end{figure*}

To quantify the speed of the diffusion in $E$ and $L$, we evaluated the spread $\sigma_t[E]$ as the difference between the third and the first quartiles of the distribution of energy of individual particles $E_i(t)$. This method is more robust to outliers than the calculation of the standard deviation $\sigma$, while it gives $1.35\,\sigma$ for a normal distribution. Similarly, $\sigma_t[L]$ was obtained for the angular momentum shifts $L_i(t)-L_i(0)$. In the diffusion driven by two-body relaxation, one would expect
\begin{equation}
    \frac{\sigma^2_t[E]}{W^2} = \zeta \frac{t}{t_{\rm relax}}\,,\quad \frac{\sigma^2_t[L]}{L^2_{\rm circ}} = \eta \frac{t}{t_{\rm relax}}\,
\end{equation}
where $\zeta$, $\eta$ are constants of order unity, $W$ is some energy characteristic of the system, e.g. total kinetic energy, or virial of Clausius $W_{\rm C}$~\citep[e.g.,][]{2015MNRAS.453.2919S}. If the resonant relaxation takes place for angular momentum, one would expect $\eta \sim (\varepsilon \ln N)^{-1}$.

The character of the diffusion can be understood by analyzing the plots summarised in Fig.\,\ref{fig:DE}. The upper left panel shows the normalized energy diffusion coefficient
\begin{equation}
    D[E] \equiv \frac{t_{\rm relax}}{W^2_{\rm C}} \frac{\d }{\d t}\sigma^2_t[E]\,,
    \label{eq:D-E}
\end{equation}
By construction, it should give the constant value $\zeta$ for two-body relaxation. In fact, we observe an almost constant value of order unity, which indicates that spreading in energy is due to two-body relaxation. In the upper right panel, we present the energy relaxation time calculated as
\begin{equation}
    T[E] \equiv \left[ \frac1{W^2_{\rm C}} \frac{\d }{\d t}\sigma^2_t[E] \right]^{-1}
    \label{eq:T-E}
\end{equation}
for nine available runs in which the central mass $M$ varied from 3 to 100, and the number of particles $N$ varied from $25\cdot 10^3$ to $10^6$. Dashed lines show fits
\begin{equation}
    t_{\rm relax.fit} = 0.43 M^{3/2} N/\ln N \,,
    \label{eq:relaxfit}
\end{equation}
for various $M$, demonstrating that the scaling perfectly corresponds to the two-body relaxation time.

The profiles for angular momentum diffusion consist of two distinctive parts: nearly constant in the beginning and modulated exponential growth and fall later. Using the first constant part, we evaluated the angular momentum relaxation time
\begin{equation}
    T[L] \equiv \left[ \frac1{L^2_{\rm circ}} \frac{\d }{\d t}\sigma^2_t[L] \right]^{-1}\,,
    \label{eq:T-L}
\end{equation}
which is presented in the lower right panel of Fig.\,\ref{fig:DE}. Color dashed lines show fits
\begin{equation}
    t_{\rm res.fit} = 0.75 M^{2/3} N/\ln N \,,
    \label{eq:resfit}
\end{equation}
which is close to the predicted scailing of resonant relaxation $ t_{\rm res} \propto M^{1/2} N$~\citep{1996NewA....1..149R}, shown by the black dashed line.

Using the dependence of Eq.\,(\ref{eq:resfit}), we construct the normalized angular momentum diffusion coefficient
\begin{equation}
    D[L] \equiv \frac{t_{\rm res.fit}}{L^2_{\rm circ}} \frac{\d }{\d t}\sigma^2_t[L]\,.
    \label{eq:D-L}
\end{equation}
The profiles are shown in the lower left panel of Fig.\,\ref{fig:DE}. Due to normalization, the constant parts are close to unity. The growing parts are obviously due to instabilty, as follows from a comparison with the theoretical slope for the octupole harmonic given in Fig.\,\ref{fig:a3abs}.

\section{Conclusions and final remarks}
\label{sec:summary}

The gravitational loss-cone instability (gLCI) in a dominating Keplerian (near-K) potential predicted earlier theoretically~\citep{2007MNRAS.379..573P, 2008MNRAS.386.1966P}, is now revealed for the first time in numerical simulations. In the limit of slow precessing orbital motion, it proved possible to connect the sign of the orbital precession rate with the derivative of the DF with respect to the angular momentum. It was found that gLCI is possible if $\p F/\p L\cdot \Omega_{\rm pr}<0$. For the dominating Keplerian potential, the precession is {\it always} retrograde ($\Omega_{\rm pr}<0$), i.e., independently of the cluster DF, so gLCI requires $\p F/\p L > 0$ at small $L$, or a deficit of particles at small $L$ (the loss cone) must be present. This instability has a well-known counterpart in plasma physics called loss-cone instability~\citep{1965PhFl....8..547R}.

Our N-body runs typically consist of $N\sim 10^5$ particles evaluated during $\lesssim 10^4$ dynamical times. The main issue is the dominating central mass which requires special treatment of the orbital integration near the center.
The simulations reproduce well the linear regime for which instability of the octupole harmonic was predicted. In particular, we obtained the growth rate of the harmonic amplitude and periodic modulations of the amplitudes due to the presence of the complementary unstable eigen-mode (i.e. the mode with the eigen-frequency $-\omega^*$). However, the shape of the cluster does not change due to early instability saturation. The latter occurs due to diffusion enhanced by the instability.

A comparison to the near-harmonic (near-h) case~\citep{P19h} shows the following distinctive features:

\noindent
--~energy diffusion is driven by two-body relaxation in the near-K case, while it is enhanced by instability in near-h case;

\noindent
--~in the near-K case, angular momentum diffusion is driven by resonant relaxation until the instability enhanced diffusion takes over;

\noindent
--~in the near-h case, angular momentum diffusion is enhanced by instability from the very beginning, but we found no signs of resonant relaxation.

In the future, we plan to explore a diversity of more realistic DFs to detect the gLCI. We also plan to modify the numerical code so that the stars diffused to the loss cone will be excluded from the simulations. This may help to sustain the instability.

\section*{Acknowledgments}
This work was supported by the Deutsche Forschungsgemeinschaft (DFG, German Research Foundation) -- Project-ID 138713538 -- SFB 881 (``The Milky Way System'', subproject A06), by the Volkswagen Foundation under the Trilateral Partnerships grants No. 90411, 97778, and the Basic Research program II.16 (Ilia Shukhman). Peter Berczik acknowledges support by the Chinese Academy of Sciences through the Silk Road Project at NAOC, through the ``Qian\-ren'' special foreign experts program, and the President’s International Fellowship for Visiting Scientists program of CAS, the National Science Foundation of China under grant No. 11673032 and also the Strategic Priority Research Program (Pilot B) ``Multi-wavelength gravitational wave universe'' of the Chinese Academy of Sciences (No. XDB23040100). The special GPU accelerated supercomputer Laohu at NAOC has been used and we thank the Center of Information and Computing of NAOC for support. Peter Berczik also acknowledges the special support by the NASU under the Main Astronomical Observatory GRID/GPU computing cluster project. This work benefited from support by the International Space Science Institute, Bern, Switzerland,  through its International Team program ref. no. 393 ``The Evolution of Rich Stellar Populations \& BH Binaries'' (2017-18).

\appendix
\section{Spherical harmonics in N-body simulations}
\label{app1}

We start with a smooth density distribution
\begin{equation}
\rho(r,\theta,\varphi)=\sum\limits_{l=0}^{\infty} \rho_l(r,\theta,\varphi)\,,
\label{eq:rho}
\end{equation}
where
\begin{equation}
\rho_l(r,\theta,\varphi)=\sum\limits_{m=-l}^l C_l^m(r)\, Y_l^m(\theta,\phi)\,,
\end{equation}
and $Y_l^m(\theta,\phi)$ are fully normalised spherical harmonics:
\begin{multline}
\int d\Omega\, Y_l^m(\theta,\varphi)[Y_{l'}^{m'}(\theta,\varphi)]^* \equiv\\
\equiv  \int\limits_0^{\pi} \sin\theta\, d\theta\int\limits_0^{2\pi} d\varphi \,Y_l^m(\theta,\varphi) [Y_{l'}^{m'}(\theta,\varphi)]^*= \delta_{l l'}\,\delta_{m m'}
\end{multline}
($[...]^*$ denotes the complex conjugate). This orthogonality gives the coefficients of the expansion:
\begin{equation}
C_l^m(r)= \int d\Omega\, \rho_l(r,\theta,\varphi)\, [Y_l^m(\theta,\varphi)]^*,\ \ |m|\le l \,.
\end{equation}
In models of clusters with a DF depending on $E$ and $L$ only, the time dependence (i.e. eigenfrequencies of oscillations) is independent of $m$~\citep[e.g.,][]{1984sv...bookQ....F}.

Now we consider a particle distribution in N-body simulations
\begin{equation}
\rho(r,\theta,\phi)= \sum\limits_{i=1}^N \mu_i \frac{\delta(r-r_i)}{r^2}\,\frac{\delta(\theta-\theta_i)}
{\sin\theta}\,\delta(\varphi-\varphi_i)
\end{equation}
(here $N$ is the total number of particles; $\mu_i$, $r_i$, $\theta_i$ and $\varphi_i$ are mass and spherical coordinates of particle $i$) and introduce a {\it global} characteristic of each harmonic component as follows:
\begin{equation}
A_l^m = \frac{1}{M_*} \int \d r  \, r^2  C_l^m(r)\, = \frac{1}{M_*}\sum\limits_{i=1}^N \mu_i [Y_l^m(\theta_i,\varphi_i)]^*\,.
\end{equation}
The strength of the spherical harmonic $l$ can be described by coefficients $A_l$, where:
\begin{equation}
A_l^2 \equiv \sum_{m=-l}^l |A_l^m|^2\,.
\label{eq:str}
\end{equation}
Using the addition theorem for spherical harmonics~\citep[e.g.][]{Arfken}:
\begin{equation}
P_l(\cos\Theta) = \frac{4\pi}{2l+1} \sum_{m=-l}^l Y_l^{m*} (\theta', \varphi') Y_l^{m} (\theta, \varphi)\,,
\end{equation}
($P_l(x)$ are the Legendre polynomials, $\Theta$ is an angle between directions $(\theta, \varphi)$ and $(\theta', \varphi')$), one can show the independence of $A_l$ (Eq.\,\ref{eq:str}) of the orientation of the coordinate frame:
\begin{align}
    N^{2}A_l^2 ={}&  N^{2} \sum_{m=-l}^l |A_l^m|^2 = \notag \\
  ={}& \sum_{m=-l}^l \sum\limits_{i=1}^N \mu_i Y_l^{m} (\theta_i, \varphi_i)
                 \sum\limits_{j=1}^N \mu_j Y_l^{m*}(\theta_j, \varphi_j) = \notag \\
  ={}& \sum\limits_{i=1}^N \sum\limits_{j=1}^N \mu_i \mu_j
                     \sum_{m=-l}^l Y_l^{m}(\theta_i, \varphi_i) Y_l^{m*}(\theta_j, \varphi_j) = \notag \\
  ={}& \frac{2l+1}{4\pi} \sum\limits_{i=1}^N \sum\limits_{j=1}^N \mu_i \mu_j P_l(\cos\Theta_{ij}) = \notag \\
  ={}& \frac{2l+1}{4\pi} \left[ \sum\limits_{i=1}^N \mu^2_i  +
                                \sum\limits_{i\ne j} \mu_i \mu_j P_l(\cos\Theta_{ij}) \right]\,
\end{align}
($\Theta_{ij}$ denotes an angle between particle $i$ and $j$). The last expression is clearly independent of the orientation of the frame. If masses of the particles are equal, $\mu_i = M_*/N$, then
\begin{equation}
A_l = \frac1N \left[ \sum_{m=-l}^l \left| \sum\limits_{i=1}^N Y_l^{m} (\theta_i, \varphi_i)\right|^2 \right]^{1/2}\,.
\label{eq:A10}
\end{equation}
This expression is used to evaluate the strength of spherical harmonics in our N-body simulations.



\end{document}